\documentclass[10pt]{article}

\usepackage{color}
\usepackage{dsfont}
\usepackage{enumerate}
\usepackage{graphicx,float}
\usepackage{caption}
\usepackage{subcaption}
\usepackage{setspace}
\usepackage{hyperref}
\usepackage{color}
\usepackage{diagbox}
\usepackage[english]{babel}
\usepackage{amsmath, amsfonts, amssymb, amsthm, amscd,graphicx}

\newcommand{\comment}[1]{}

\newcommand{\bee}{\begin{equation}}
\newcommand{\eee}{\end{equation}}
\newcommand{\bea}{\begin{eqnarray}}
\newcommand{\eea}{\end{eqnarray}}
\newcommand{\bean}{\begin{eqnarray*}}
\newcommand{\eean}{\end{eqnarray*}}

\begin{document}


\title{The quadratic rough Heston  model and \\the joint S\&P $500$/VIX smile calibration problem}

		\author{Jim Gatheral\footnote{Baruch College, CUNY,  jim.gatheral@baruch.cuny.edu},~~ 
		Paul Jusselin\footnote{\'Ecole Polytechnique, paul.jusselin@polytechnique.edu}~~and Mathieu Rosenbaum\footnote{\'Ecole Polytechnique, mathieu.rosenbaum@polytechnique.edu} \\
		}
\date{\today }
\maketitle

\begin{abstract}
\noindent Fitting simultaneously SPX and VIX smiles is known to be one of the most challenging problems in volatility modeling. A long-standing conjecture due to Julien Guyon is that it may not be possible to calibrate jointly these two quantities with a model with continuous sample-paths. 
We present the quadratic rough Heston  model as a counterexample to this conjecture.  The key idea is the combination of rough volatility together with a price-feedback (Zumbach) effect.
\end{abstract}


\section{Introduction}
\label{sec:introduction}

The Volatility Index, or VIX, was introduced in 1993 by the Chicago Board Options Exchange (CBOE for short) and was originally designed according to \cite{vixwhite} to ``measure the market’s expectation of 30-day volatility implied by at-the-money S\&P 100 Index option price''. Since 2003, the VIX has been redefined as the square root of the price of a specific basket of options on the S\&P 500 Index (SPX) with maturity $30$ days. The basket coefficients are chosen so that at any time $t$, the VIX represents the annualized square root of the price of a contract with payoff equal to $\log(S_{t+\Delta}/S_t)$ where $\Delta = 30$ days and $S$ denotes the value of the SPX. Consequently, it can be formally written via risk-neutral expectation under the form
\begin{equation}
\label{eq:vix_def}
\text{VIX}_t = \sqrt{\mathbb{E}[\log(S_{t+\Delta}/S_t)|\mathcal{F}_t]} \times 100,
\end{equation}
where $(\mathcal{F}_t)_{t\geq 0}$ is the natural filtration of the market.\\

\noindent Since 2004, investors have been able to trade VIX futures. Quoting the CBOE white paper, they ``provide market participants with a variety of opportunities to implement their view using volatility trading strategies, including risk management, alpha generation and portfolio diversification''\footnote{\url{https://cfe.cboe.com/cfe-products/vx-cboe-volatility-index-vix-futures}}. Subsequently in 2006, CBOE introduced VIX options ``providing market participants with another tool to manage volatility. VIX options enable market participants to hedge portfolio volatility risk distinct from market price risk and trade based on their view of the future direction or movement of volatility''\footnote{\url{http://www.cboe.com/products/vix-index-volatility/vix-options-and-futures/vix-options}}. Those products are now among the most liquid financial instruments in the world. There are indeed more than 500,000 VIX options traded each day, with most of the liquidity concentrated on the first three monthly contracts.\\

\noindent Nevertheless, despite that more vega is now traded in the VIX  market than in the SPX market, the wide bid-ask spreads in the VIX options market betray its lack of maturity. One of the reasons underlying these wide spreads is that the market lacks a reliable pricing methodology for VIX options; since the VIX is by definition a derivative of the SPX, any reasonable methodology must necessarily be consistent with the pricing of SPX options. Designing a model that jointly calibrates SPX and VIX options prices is known to be extremely challenging. Indeed, this problem is sometimes considered to be {\it the holy grail of volatility modeling}. We will simply refer to it as the \textit{joint calibration problem.}\\

The joint calibration problem has been extensively studied by Julien Guyon who provides a review of various approaches in \cite{guyon2019joint}. We can split the different attempts to solve it into three categories. In probably the most technical and original proposal, and the first to have succeeded in obtaining a perfect joint calibration, the joint calibration problem is interpreted as a model-free constrained martingale transport problem, as initially observed in \cite{de2015linking}. In his recent paper \cite{guyon2019joint}, using this viewpoint, Guyon manages to get a perfect calibration of VIX options smile at time $T_1$ and SPX options smiles at dates $T_1$ and $T_2 = T_1+30$ days. As noticed by the author, although this methodology can theoretically be extended to any set of maturities, it is much more intricate in practice because of the computational complexity.\\

This drawback is avoided in the second and third types of approach where models are in continuous-time. Continuous-time models have the advantage that they rely on observable properties of assets and so allow for practical intuition on their dynamics. The second approach is to attempt joint calibration with models where SPX trajectories are continuous, see in particular \cite{goutte2017regime}.  Unfortunately, for now, continuous models have not been completely successful in this task. An interpretation for this failure is given in \cite{guyon2019joint} where the author explains that ``the very large negative skew of short-term SPX options, which in continuous models implies a very large volatility of volatility, seems inconsistent with the comparatively low levels of VIX implied volatilities''. To circumvent this issue, it is then natural to think of rough volatility models as recently introduced in \cite{gatheral2018volatility}. However, these models also appear unsuccessful thus far, see \cite{guyon2018joint}.\\

\noindent The last approach is to allow for jumps in the dynamic of the SPX, see \cite{baldeaux2014consistent, cont2013consistent, kokholm2015joint, pacati2018smiling, papanicolaou2014regime}. Doing so, one can reconcile the skewness of SPX options with the level of VIX implied volatilities. Nevertheless, probably besides those in \cite{cont2013consistent} and \cite{pacati2018smiling}, existing models with jumps do not really achieve a satisfying accuracy for the joint calibration problem. Specifically, most of them fail to reproduce VIX smiles for maturities shorter than one month.\\

\noindent In summary, according to Guyon in \cite{guyon2019joint}, despite the many efforts ``so far all the attempts at solving the joint SPX/VIX smile calibration problem [using a continuous time model] only produced imperfect, approximate fits''. In particular, regarding continuous models, Guyon concludes that ``joint calibration seems out of the reach of continuous-time models with continuous SPX paths''. In this paper, we provide a counterexample to  Guyon's conjecture, namely a model with continuous SPX and VIX paths that enables us to fit SPX and VIX options smiles simultaneously.


\section{Rough volatility and the Zumbach effect}


Recently rough volatility models, where volatility trajectories, though continuous, are very irregular, have generated a lot of attention. The reason for this success is the ability of these very parsimonious models to reproduce all the main stylized facts of historical volatility time series and to fit SPX options smiles, see \cite{bayer2016pricing,euch2019roughening, gatheral2018volatility}. One particularly interesting rough volatility model is the rough Heston model introduced in \cite{el2019characteristic} which as its name suggests, is a rough version of the classical Heston model. This model arises as limit of natural Hawkes process-based models of price and order flow, see \cite{el2018microstructural, jaisson2016rough, jusselin2018no}. Moreover, there is a quasi-closed form formula for the characteristic function of the rough Heston model, just as in the classical case. So fast pricing of European options is possible, see \cite{callegaro2018rough, gatheral2019rational}. In addition derivatives hedging is fully understood as shown in \cite{el2018perfect, el2019characteristic}, see also \cite{jaber2019affine, cuchiero2018generalized}. \\

Despite these successes, a subtle question raised by Jean-Philippe Bouchaud remains: can a rough volatility model  reproduce the so-called {\it Zumbach effect}, the observation originally due to Gilles Zumbach, see \cite{lynch2003market,zumbach2009time, zumbach2010volatility}, that financial time series are not time-reversal invariant?  To answer this question, we introduce two notions, each of which corresponds to different aspects of the Zumbach effect: 

\begin{itemize}

\item[-]The {\it weak Zumbach effect} (typically considered in the econophysics literature, see \cite{zumbach2009time}): Past squared returns forecast better future integrated volatilities than past integrated volatilities forecast future squared returns. This property is not satisfied in classical stochastic volatility models. However, rough stochastic volatility models are consistent with the weak Zumbach effect, see \cite{el2019zumbach} for explicit computations in the rough Heston model.

\item[-]The {\it strong Zumbach effect}: Conditional dynamics of volatility with respect to the past depend not only on the past volatility trajectory but also on the historical price path; specifically, price trends tend to increase volatility, see \cite{zumbach2010volatility}. Such feedback of the historical price path on volatility also occurs on implied volatility as illustrated in Figure 1 of \cite{foschi2008path} and in \cite{zumbach2010volatility}. Rough stochastic volatility models such as the rough Heston model are not consistent with the strong Zumbach effect, see \cite{el2018perfect}.

\end{itemize} 

The quest for a rough volatility model consistent with the strong Zumbach effect and the empirical success of quadratic Hawkes process-based models documented in \cite{blanc2017quadratic} led to the development of super-Heston rough volatility models in \cite{dandapani2019quadratic}.  These extensions of the rough Heston model arise as limits of quadratic Hawkes process-based microstructural models just as the rough Heston model arises as the continuous-time limit of a linear Hawkes process-based microstructural model.\\

The idea of using super-Heston rough volatility models to solve the joint calibration problem came after a presentation of Julien Guyon at \'Ecole Polytechnique in March 2019. In this talk, he gave a necessary condition for a continuous model to fit simultaneously SPX and VIX smiles: The inversion of convex ordering between volatility and the local volatility implied by option prices, see \cite{acciaio2019inversion, guyon2019inversion}. The intuition behind this condition could be reinterpreted as some kind of strong Zumbach effect. It was then natural for us to investigate the ability of super-Heston rough volatility models to solve the joint calibration problem.


\section{The quadratic rough Heston  model}
\label{sec:model_definition}

%
%
%
The quadratic rough Heston  model that we consider is a special case of the super-Heston rough volatility models of \cite{dandapani2019quadratic}. 
The joint dynamics of the asset $S$ (here the SPX), and its spot variance $V$ satisfy
\begin{equation*}
\mathrm{d}S_t=S_t \sqrt{V_t}dW_t,~  V_t = a(Z_t - b)^2 +c,
\end{equation*}
where $W$ is a Brownian motion, $a$, $b$ and $c$ some positive constants and $Z_t$ follows a rough Heston model. More precisely,
\begin{equation*}
Z_t = \int_0^t (t-s)^{\alpha-1} \frac{\lambda}{\Gamma(\alpha)} (\theta_0(s) - Z_s) \mathrm{d}s + \int_0^t (t-s)^{\alpha-1} \frac{\lambda}{\Gamma(\alpha)}  \eta \sqrt{V_s} \mathrm{d}W_s,
\end{equation*}
with $\alpha \in (1/2,1)$, $\lambda>0$, $\eta>0$ and $\theta_0$ a deterministic function.
In this special case of a quadratic rough Heston model, the asset $S$ and its volatility depend on the history of only one Brownian motion.  The model is thus a pure feedback model;  volatility is driven only by the price dynamics, with no additional source of randomness.   In general of course, the volatility process does not need to depend only on the Brownian motion driving the asset price $S$.  For simplicity however, we will refer to this pure feedback version of the quadratic rough Heston model as the {\em quadratic rough Heston model}.

\subsection{The quadratic rough Heston  process}


The process $Z_t$ may be understood as a weighted moving average of past price log returns. Indeed from Lemma A.1 in \cite{el2018perfect}, we have that 
$$
Z_t=\int_0^t f^{\alpha, \lambda}(t-s) \theta_0(s)\mathrm{d}s + \int_0^t f^{\alpha, \lambda}(t-s) \eta \sqrt{V_s}\mathrm{d}W_s,
$$
where $f^{\alpha, \lambda}(t)$ is the Mittag-Leffler density function defined for $t\geq 0$ as
$$
f^{\alpha, \lambda}(t)=\lambda t^{\alpha-1} E_{\alpha, \alpha}(-\lambda t^{\alpha}),$$ with
$$E_{\alpha, \beta}(z) = \sum_{n\geq 0}\frac{z^n}{\Gamma(\alpha n+\beta)}.
$$
The variable $Z_t$ is therefore {\it path-dependent}, a weighted average of past returns of the type typically considered in path-dependent volatility models, see \cite{hobson1998complete}.  As explained in \cite{guyon2014path}, modeling with path-dependent variables is a natural way to reproduce the fact that volatility depends on recent price changes. However the kernels used to model this dependency are typically exponential, see for example \cite{hobson1998complete}. Here a crucial idea, motivated by our previous work \cite{dandapani2019quadratic}, is to use a rough kernel, more precisely the Mittag-Leffler density function. Thanks to this kernel, the ``memory'' of $Z$ decays as a power law and $Z$ is highly sensitive to recent returns since
$$
f^{\alpha, \lambda}(t) \underset{t\rightarrow +\infty}{\sim} \frac{\alpha}{\lambda \Gamma(1-\alpha)}t^{-\alpha - 1} \text{ and } f^{\alpha, \lambda}(t) \underset{t\rightarrow 0^+}{\sim} \frac{ \lambda  }{\Gamma(\alpha)}t^{\alpha - 1}.
$$
This essentially means that long periods of trends or sudden upwards or downwards moves of the price generate large values for $|Z|$ and so high volatility, in particular when $Z$ is negative. Such link is clearly observed on data, see Figure \ref{fig:historical_spx_vix} where the VIX index spikes almost instantaneously after large negative returns of the SPX and then decreases slowly afterwards. We plot in Figure \ref{fig:simulation_spx_vix} an example of sample paths of SPX and VIX indexes in our model. The feedback of negative price trends on volatility is very well reproduced. Finally the choice of $f^{\alpha, \lambda}$ as kernel ensures that the volatility process is rough, with Hurst parameter equal to $H = \alpha - 1/2$. As shown in \cite{gatheral2018volatility}, this enables us to reproduce the behavior of historical volatility time series provided $H$ is taken of order $0.1$.\\

As explained above, an immediate consequence of the feedback effect is that negative price trends generate high volatility levels. But such trends also impact the instantaneous variance of volatility in our model. To see this, consider the classical case with $\alpha = 1$. In that case, an application of It\^o's Formula gives that up to a drift term,
$$
\mathrm{d}V_t = 2a (Z_t-b) \lambda \eta \sqrt{V_t} \mathrm{d}W_t.
$$
Thus the ``variance of instantaneous variance" coefficient is proportional to $ a (Z_t - b)^2$ which, up to $c$, is equal to the variance of $\log S$. Thus when volatility is high, volatility of volatility is also high.  In particular, conditional on a large downwards move in SPX, we would expect $V$ to be high and so also the volatility of $V$.  This explains why our model generates upward sloping VIX smiles.\\

We remark that incorporating the influence of price trends on volatility and instantaneous variance of volatility is the main motivation underlying the model of \cite{goutte2017regime}. That model, although not solving the joint calibration problem, is probably the best of the continuous models introduced so far. In this switching model, the price follows a classical Heston dynamic where the parameters can change depending on the value of an hidden Markov chain with three states. It is motivated by a $100$-days rolling calibration of the classical Heston model performed by the authors, see Figure 2 in \cite{goutte2017regime}. This rolling calibration suggests that volatility, leverage and volatility of volatility are higher in period of crisis. Hence they introduce a Markov chain to trigger crisis phases and switch the parameters of the Heston model depending on the situation. The three possible states of the chain can therefore be interpreted as corresponding to the following situations: 
\begin{itemize}
\item[-] Flat or increasing SPX.
\item[-] Transition phase between flat SPX and crisis.
\item[-] Crisis with dramatically decreasing SPX.
\end{itemize}
The Markov chain in \cite{goutte2017regime} can therefore somehow be seen as an {\it ad hoc} version of the process $Z$ in the quadratic rough Heston  model.

\subsection{Parameter interpretation}

\begin{figure}[h!]
\begin{center}
\includegraphics[width=16cm,height=8cm]{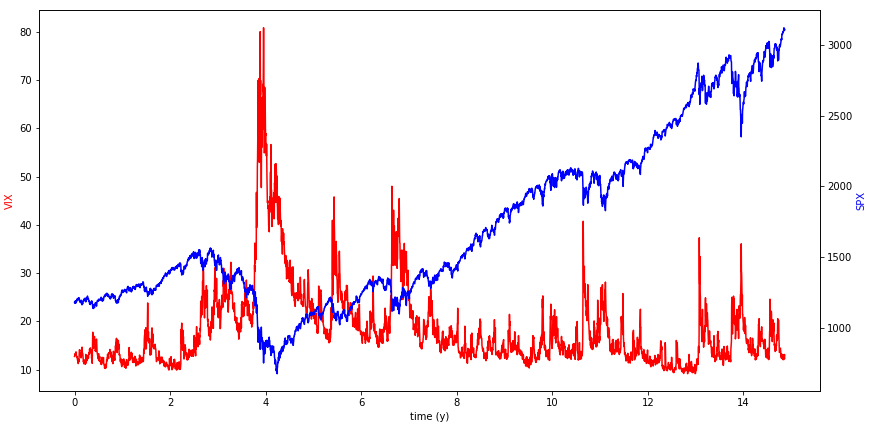}
\caption{SPX (in blue) and VIX (in red) indexes from $25$ November 2004 to $25$ November 2019.}
\label{fig:historical_spx_vix}
\end{center}
\end{figure}

\begin{figure}[h!]
\begin{center}
\includegraphics[width=16cm,height=8cm]{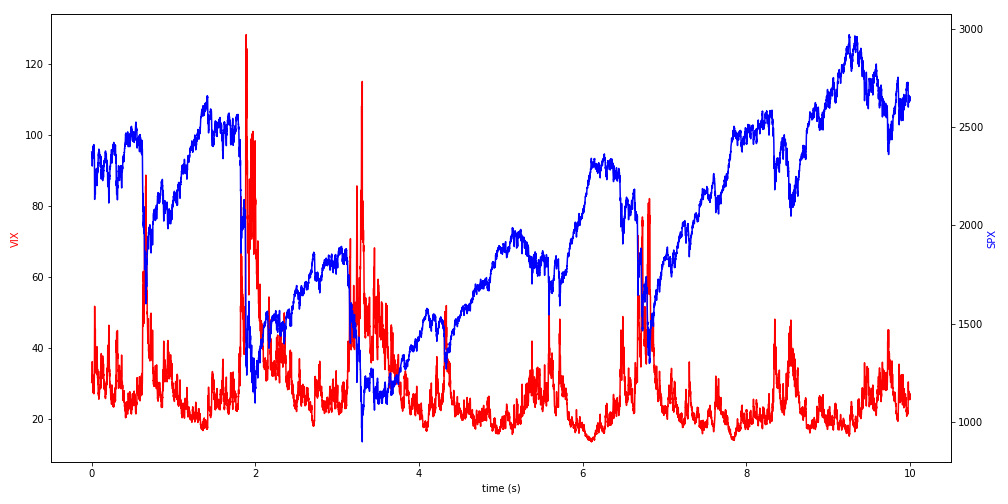}
\caption{SPX (in blue) and VIX (in red) indexes from simulation of the quadratic rough Heston model.}
\label{fig:simulation_spx_vix}
\end{center}
\end{figure}

The parameters $a$, $b$ and $c$ in the specification
\[
V_t = a(Z_t - b)^2 + c
\]
can be interpreted in the following way.
\begin{itemize}
\item[-] $c$ represents the minimal instantaneous variance. When calibrating the model, we use $c$ to shift upward or downward the smiles of SPX options.

\item[-] $b>0$ encodes the asymmetry of the feedback effect. Indeed for the same absolute value of $Z$, volatility is higher when $Z$ is negative than when it is positive. Such asymmetry aims at reproducing the empirical behavior of the VIX. This is illustrated in Figure \ref{fig:historical_spx_vix} where we observe that the VIX spikes when the SPX tumbles down, but not after it goes up. From a calibration viewpoint, the higher $b$ the more SPX options smiles are shifted to the right. 

\item[-]  $a$ is the sensitivity of the volatility to the feedback of price returns. The greater $a$, the greater the role of  feedback in the model and the higher is volatility of volatility. Consistent with this SPX smiles become more extreme as $a$ increases.
\end{itemize}

\subsection{Infinite dimensional Markovian representation}

Though the quadratic rough Heston  model is not Markovian in the variables $(S, V)$, it does admit an infinite dimensional Markovian representation. Inspired by the computations in \cite{el2018perfect}, we obtain that for any $t$ and $t_0$ positive
\begin{equation}
\label{eq:fwd_vol}
Z_{t_0+t} = \int_0^t (t-s)^{\alpha-1} \frac{\lambda}{\Gamma(\alpha)} (\theta_{t_0}(s) - Z_{t_0+s}) \mathrm{d}s + \int_0^t (t-s)^{\alpha-1} \frac{\lambda}{\Gamma(\alpha)}  \eta \sqrt{V_{t_0+s}} \mathrm{d}W_{t_0+s},
\end{equation}
with $\theta_{t_0}$ a $\mathcal{F}_{t_0}$-measurable function. More precisely $\theta_{t_0}$ is given by
\begin{equation*}
\theta_{t_0}(u) = \theta_0(t+u) + \frac{\alpha}{\lambda\Gamma(1-\alpha)}\int_0^{t_0} (t_0-v+u)^{-1-\alpha}(Z_v - Z_{t_0})\mathrm{d}v.
\end{equation*}
Equation \eqref{eq:fwd_vol} implies that the law of $(S_t, V_t)_{t\geq t_0}$ only depends on $S_{t_0}$ and $\theta_{t_0}$. In view of \eqref{eq:vix_def} and using the same methodology as in \cite{el2018perfect}, it means that we can express the VIX at time $t$ as a function of $\theta_t$ and $S_t$. Consequently we can write the price of any European option with pay-off depending on SPX and VIX as a function of time, $S$ and $\theta$.

\section{Numerical results}
\label{sec:numerical results}

In this section, we illustrate how successfully we can fit both SPX and VIX smiles on May 19, 2017\footnote{Market data is from OptionMetrics via Wharton Data Research Services (WRDS).}, one of the dates considered in \cite{euch2019roughening}, an otherwise randomly chosen date.
We focus on short expirations, from $2$ to $5$ weeks, where the bulk of VIX liquidity is.  Moreover, short-dated smiles are the ones that are typically poorly fitted by conventional models.\\

In the quadratic rough Heston model, the function $\theta_0(\cdot)$ needs to be calibrated to market data. In the rough Heston model there is a simple bijection between $\theta_0(\cdot)$ and the forward variance curve. In the quadratic rough Heston  model, this connection is more intricate and so for simplicity we choose the following restrictive parametric form for $Z$:
$$
Z_t = Z_0 - \int_0^t (t-s)^{\alpha - 1} \frac{\lambda}{\Gamma(\alpha)} Z_s \mathrm{d}s + \int_0^t (t-s)^{\alpha - 1} \frac{\lambda}{\Gamma(\alpha)} \eta \sqrt{V_s} \mathrm{d}W_s,
$$
which is equivalent to taking \[\theta_0(t) = \frac{Z_0}{\lambda \Gamma(1-\alpha) } t^{-\alpha}.\]
Allowing $\theta_0(\cdot)$ to belong to a larger space would obviously lead to even better results, but would require a more complex calibration methodology.
Thus we are left to calibrate the parameters $\nu = (\alpha, \lambda, a, b, c, Z_0)$. We use the following objective function:$$F(\nu) = \frac{1}{\# \mathcal{O}^{SPX}}\sum_{o \in \mathcal{O}^{SPX}} (\sigma^{o,mid} - \sigma^{o, \nu})^2 + \frac{1}{\# \mathcal{O}^{VIX}}\sum_{o\in \mathcal{O}^{VIX}}(\sigma^{o,mid} - \sigma^{o, \nu})^2,
$$
where $\mathcal{O}^{SPX}$ is the set of SPX options, $\mathcal{O}^{VIX}$ the set of VIX options, $\sigma^{o, mid}$ denotes the market mid implied volatility for the option $o$ and $\sigma^{o,\nu}$ is the implied volatility of the option $o$ in the quadratic rough Heston model with parameter $\nu$ obtained by Monte-Carlo simulations. To calibrate the model, we minimize the function $F$ over a grid centered around an initial guess $\nu_0$.\\

\begin{figure}[tbph!]
\begin{center}
\includegraphics[width=16cm,height=8cm]{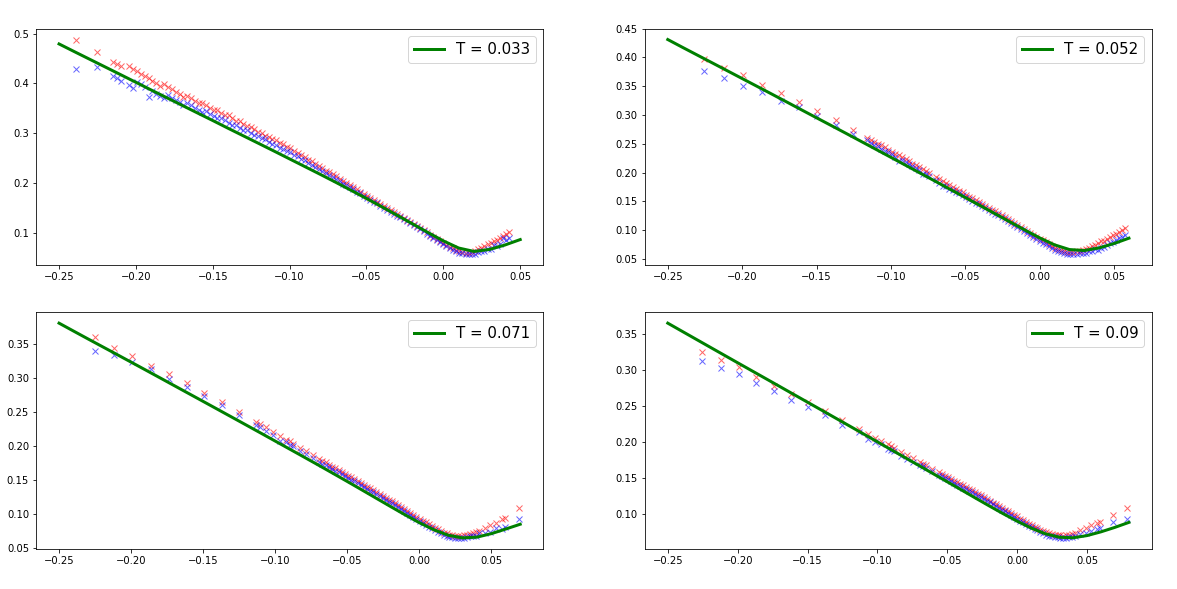}
\caption{Implied volatility on SPX options for $19$ May 2017. The blue and red points are respectively the bid and ask of market implied volatilities. The implied volatility smiles from the model are in green. The strikes are in log-moneyness and $T$ is time to expiry in years.}
\label{fig:vol_spx}
\end{center}
\end{figure}

\begin{figure}[tbph!]
\begin{center}
\includegraphics[width=16cm,height=8cm]{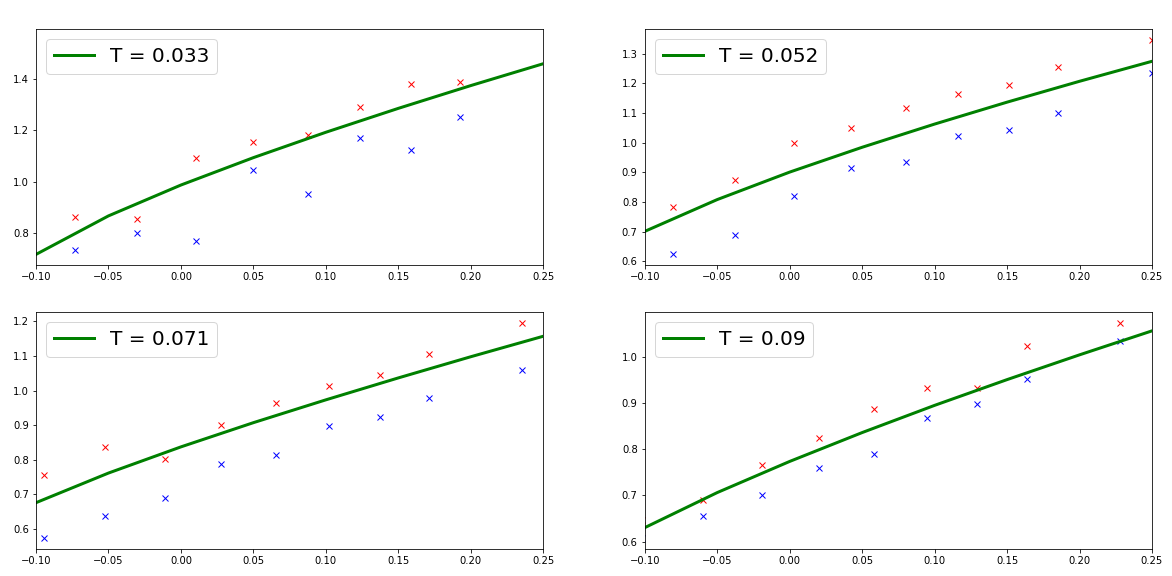}
\caption{Implied volatility on VIX options for $19$ May 2017. The blue and red points are respectively the bid and ask of market implied volatilities. The implied volatility smiles from the model are in green. The strikes are in log-moneyness and $T$ is  time to expiry in years.}
\label{fig:vol_vix}
\end{center}
\end{figure}

\noindent We obtain the following parameters:
\begin{equation}
\alpha = 0.51;~ \lambda = 1.2;~a = 0.384;~b = 0.095;~c = 0.0025,~Z_0 = 0.1.
\label{eq:params}
\end{equation}
The corresponding SPX and VIX options smiles are plotted in Figures \ref{fig:vol_spx} and \ref{fig:vol_vix}.\\

Despite that our calibration methodology is suboptimal and we only have six parameters, VIX smiles generated by the model with parameters \eqref{eq:params} fall systematically within market bid-ask spreads. The whole SPX volatility surface is also very well reproduced, in particular the overall shape of the SPX smiles, including extreme left tails. Obviously fits can be made even greater by reducing the range of strikes of interest or by fine tuning the calibration, notably through improving the $\theta_0(\cdot)$ function. We are currently working on a fast calibration approach, inspired by recent works on neural networks, see for example \cite{horvath2019deep}.

\section*{Acknowledgments}
We thank Julien Guyon for numerous inspiring discussions and Stefano de Marco for relevant comments. Paul Jusselin and Mathieu Rosenbaum gratefully acknowledge the financial support of the {\it ERC Grant 679836 Staqamof} and of the chair {\it Analytics and Models for Regulation}.

\bibliographystyle{abbrv}
\bibliography{biblio20200106}

\end{document}